\documentclass[apex,preprint]{jjap3}
\usepackage{newtxtext,newtxmath}
\usepackage{graphicx}

\title{Twisted bilayer graphene fabricated by direct bonding in a high vacuum}
\author{Hitoshi Imamura$^{1}$, Anton Visikovskiy$^{1}$, Ryosuke Uotani$^{1}$, Takashi Kajiwara$^{1}$, Hiroshi Ando$^{2}$, Takushi Iimori$^{2}$, Kota Iwata$^{2}$, Toshio Miyamachi$^{2}$, Kan Nakatsuji$^{3}$, Kazuhiko Mase$^{4,5}$,Tetsuroh Shirasawa$^{6}$, Fumio Komori$^{2}$, Satoru Tanaka$^{1}$ \thanks{E-mail: stanaka@nucl.kyushu-u.ac.jp}}
\inst{
$^{1}$Department of Applied Quantum Physics and Nuclear Engineering, Kyushu University, Fukuoka 819-0395, Japan \\ $^{2}$The Institute for Solid State Physics, The University of Tokyo, Kashiwa 277-8581, Japan \\ $^{3}$Department of Materials Science and Engineering, Tokyo Institute of Technology, Yokohama 226-8502, Japan \\ $^ {4}$Institute of Materials Structure Science, High Energy Accelerator Research Organization (KEK), Tsukuba 305-0044, Japan \\ 
$^{5}$SOKENDAI (The Graduate University for Advanced Studies), Tsukuba 305-0801, Japan \\
$^{6}$National Metrology Institute of Japan, National Institute of Advanced Industrial Science and Technology, Tsukuba 305-0395, Japan} 

\abst{Twisted bilayer graphene (TBG), in which two monolayer graphene are stacked with an in-plane rotation angle, has recently become a hot topic due to unique electronic structures. TBG is normally produced in air by the tear-and-stack method of mechanical exfoliation and transferring graphene flakes, by which a sizable, millimeter-order area, and importantly clean interface between layers are hard to obtain. In this study, we resolved these problems by directly transferring the easy-to-exfoliate CVD-grown graphene on SiC substrate to graphene in a high vacuum without using any transfer assisting medium and observed electronic band modulations due to the strong interlayer coupling.}

\begin{document}
\maketitle

Twisted bilayer graphene (TBG) is a system of two stacked and mutually in-plane rotated graphene sheets, which exhibits electronic states very different from those usually observed in typical bilayer graphene with Bernal stacking. In particular, it was shown that the Fermi velocity in the vicinity of the Dirac point strongly depends on the twist-angle below $\sim10^{\circ}$,\cite{LopesDosSantos2007, Shallcross2010, DeTramblyLaissardiere2010} and vanishes completely at the so-called magic angle of $\sim1.1^{\circ}$ resulting in a completely flat band with an extremely sharp density of states feature\cite{Bistritzer2011}. It has been predicted that these peculiar properties of electronic structures such as van Hove singularities and flat bands may lead to novel electronic characteristics.\cite{Bistritzer2011, Moon2012, TramblyDeLaissardiere2012} Indeed, the superconductivity of $\sim1.1^{\circ}$ TBG was recently experimentally observed\cite{Cao2018b} and became a hot topic. Experimental observation of flat band and other relevant electronic states is a very important issue to elucidate the mechanism of emerging superconductivity and verify theoretical propositions and calculations. Up to now most of the calculations are based on rather simplified tight-binding models\cite{Morell2010, Moon2012, Sboychakov2015} due to the enormous size of TBG periodic structure, thus verification of the results and refining of models is a pressing issue.  Recently, the electronic states of $1.34^{\circ}$ and $0.96^{\circ}$ TBG near the magic-angle have been directly evaluated by nano-angle-resolved photoemission spectroscopy (nano-ARPES) measurements, and the flat band located at the Fermi energy (E$_{\rm F}$) has actually been observed.\cite{Lisi2020b, Utama2019} Although these reports exhibited direct evidence of the flat band formation, there is still some room for discussion on electronic structure which is strongly influenced by the quality of interface/surface.\\
TBG is mostly produced by mechanical exfoliation and transfer of graphene flakes,\cite{Kim2014, Rode2016, Cao2016, Cao2018a, Cao2018b, Kerelsky2019} which essentially contaminates the interface and surface, and also restricts the size of the sample to the order of $\mu$m. The latter should especially limit the further exploration of unique TBG characteristics by the standard surface analysis techniques such as low-energy electron diffraction (LEED), reflection high-energy electron diffraction (RHEED), ARPES, surface X-ray diffraction, etc.
We, therefore, developed a method of directly bonding graphene sheets in a high vacuum without using any chemical adhesion and transmission media such as Polymethyl methacrylate (PMMA)/Polydimethylsiloxane (PDMS).\cite{Kim2016} This is an entirely new technique for simultaneously achieving large areas and a very clean interface of TBG. This technique essentially relies on two significant requirements: the growth of easy-to-exfoliate monolayer graphene on SiC and a high vacuum environment. The former is performed by our oxygen-added chemical vapor deposition (CVD) method, resulting in a new buffer layer. As a result, we achieved the millimeter size TBGs with a clean interface/surface and measured the structural parameters and electronic states using macro probe techniques.

\begin{figure}[tb]
\centering
\includegraphics[width=0.9\textwidth]{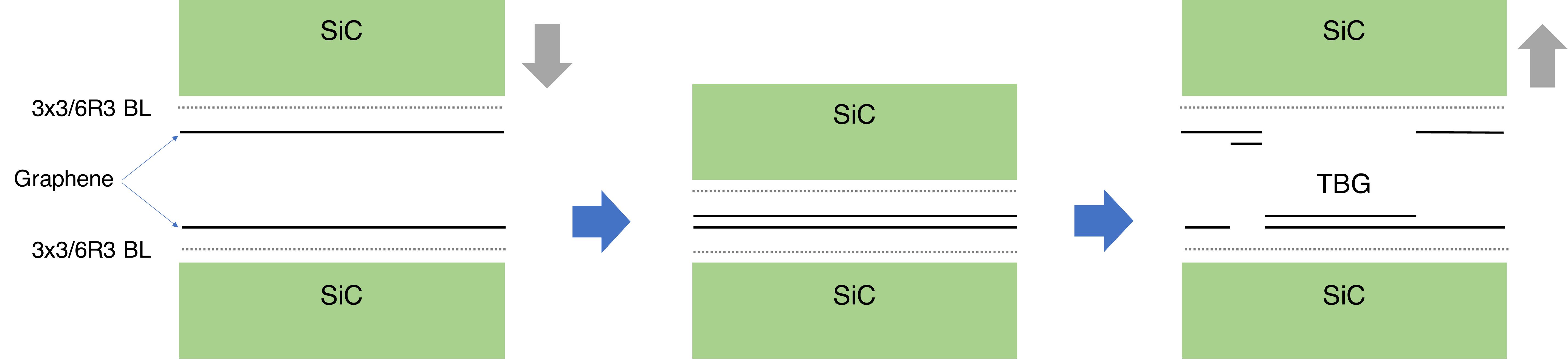}
\caption{Schematic drawings of the TBG fabrication process. (a) Two CVD grown monolayer graphene samples are set at upper and lower positions facing each other in the high vacuum chamber. (b) After annealing at $\sim200 ^{\circ}$C the samples are pressed each other at a constant pressure for 1~hour and (c) detached at the same temperature.}
\label{f1}
\end{figure}

\begin{figure}[tb]
\centering
\includegraphics[width=1.0\textwidth]{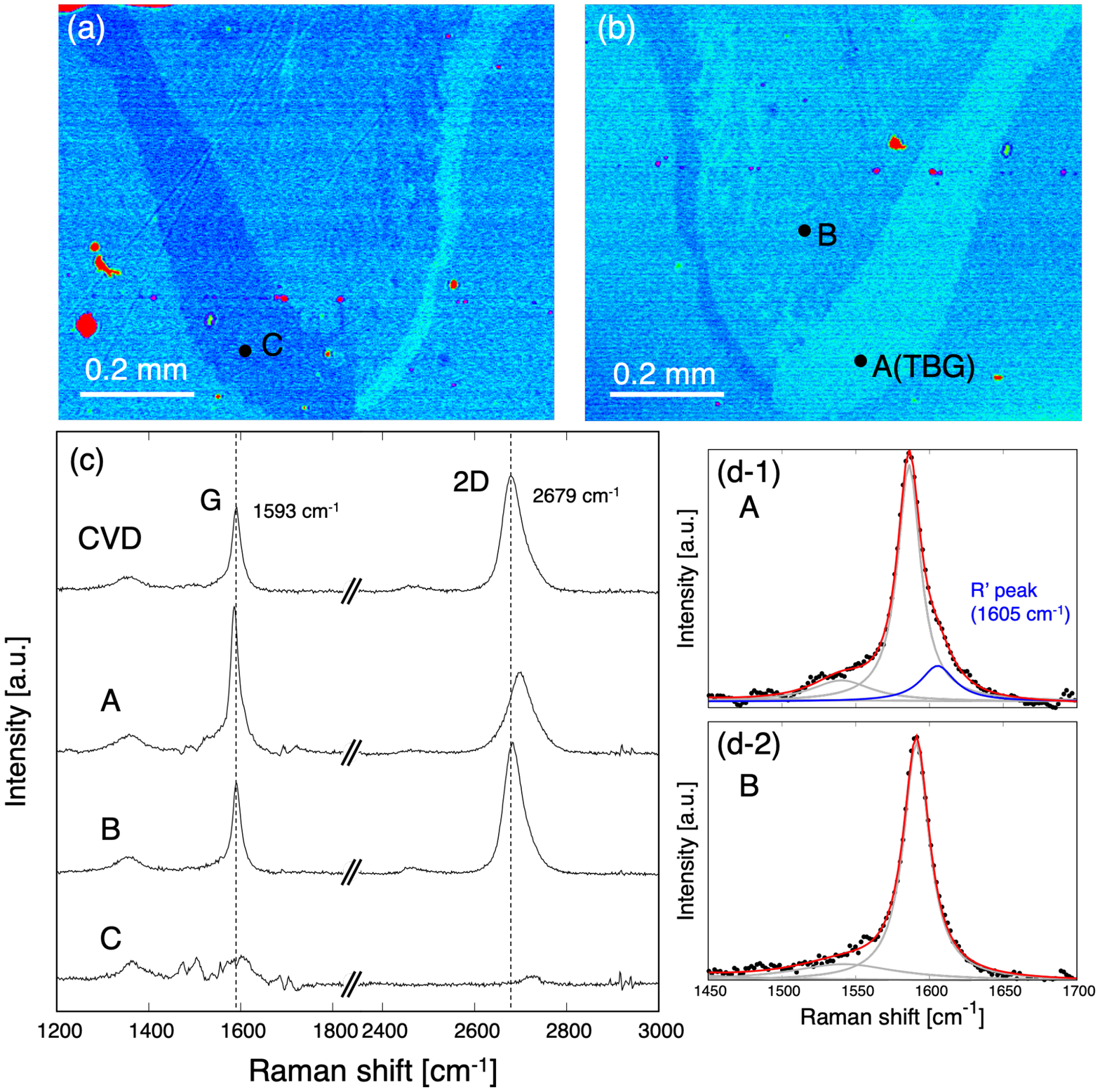}
\caption{(a), (b) Optical microscope images from the lower and upper samples at the same contact area. Three areas of different contrast are seen, corresponding to bilayer, monolayer, and no graphene regions. Letters(A, B, C) indicate the positions measured by $\mu$-Raman spectroscopy. (c) $\mu$-Raman spectra at each position and CVD-grown monolayer graphene. (d-1, d-2) G-band spectra and separated peaks of A(TBG) and B(monolayer graphene). An R$^\prime$-band is observed by the peak separation in TBG.}
\label{f2}
\end{figure}

On-axis 4H-SiC (0001) was initially etched in a hydrogen atmosphere at 1360 $^{\circ}$C to atomically smooth the surface.\cite{Owman1996} Monolayer graphene was then grown \textit{in-situ} at 1400 $^{\circ}$C by CVD using 20~ppm ethylene and $0.6-0.7$~ppm oxygen containing Ar gas at 1 atm on the SiC substrate. The samples were examined by atomic force microscopy (AFM), $\mu$-Raman spectroscopy, LEED, X-ray photoemission spectroscopy (XPS), ARPES, and X-ray crystal truncation rod (X-CTR) scattering.\cite{Takahashi} The interface structure by our CVD is different from the one obtained by thermal decomposition of SiC, which contains graphene-like $(6\sqrt3\times6\sqrt3)R30^{\circ}$ (further 6R3) buffer layer.\cite{Riedl2007} The new buffer layer composed of Si, C, and O atoms, has a $(3\times3)$ periodicity relative to SiC(0001) surface, and results in quasi-free standing monolayer graphene on-top  as evidenced by Raman 2D-band peak position at $\sim2679\textrm{~cm}^{-1}$ (see Fig.~\ref{f2}(c)). The detailed result on the interface structure will be reported elsewhere. This buffer layer allows easy exfoliation of the graphene layer from SiC. It is found that graphene can be actually exfoliated even using a sticky-tape unlike one on the 6R3 buffer, where proper metal deposition followed by complicated processing is required.\cite{Kim2014} It should be pointed that, however, there are some minor 6R3 regions still remains in the vicinity of steps in the present system. Afterwords, as schematically shown in Fig.~\ref{f1}(a), two CVD grown samples were mounted on the upper and lower susceptors facing each other in the high vacuum chamber ($\sim10^{-4}$~Pa) and set at the desired relative orientation (0$^{\circ}$ twist-angle in this case) using RHEED for referencing. After annealing at 200$^{\circ}$C for $\sim1$~hour, the samples were pressed together at a constant pressure for 1 hour (Fig.~\ref{f1}(b)) and detached at the same temperature (Fig.~\ref{f1}(c)). The samples were then evaluated by optical microscope, $\mu$-Raman spectroscopy, LEED, and ARPES.

Figure~\ref{f2}(a) and (b) show optical microscope images of corresponding same contact area from both (upper and lower) samples after detachment. Contrasted features appear indeed mirror-like, indicating that exfoliation and transfer of monolayer graphene took place. Three different contrast areas are visible: light, medium, and dark. By $\mu$-Raman spectroscopy, each region is found to correspond to bilayer, monolayer, and no graphene, respectively as indicated in Fig.~\ref{f2}(c). Spectrum B of the medium contrast region shows monolayer graphene features: high 2D/G intensity ratio and narrow 2D-band which is fitted with single Lorentzian function. These are identical to CVD-grown graphene. Thus, no exfoliation and transfer occurred in this region. On the other hand, spectra A and C from correlated areas of upper and lower samples indicate clear changes. Spectrum C shows significant decrease of G and 2D peak intensity, suggesting monolayer graphene was removed from this region. 6R3 features around G-band and small 2D peak is due to monolayer graphene on remaining 6R3 buffer layer\cite{Ni2008} at the step region which is hard to exfoliate. The exfoliated monolayer graphene was therefore transferred onto opposite surface, which is A area of light contrast. Conclusively, the light area in Fig.~\ref{f2}(a) and (b) is TBG. The TBG area in Fig.~\ref{f2}(b) is 0.2~mm $\times$ 1.0~mm in size. The Raman spectrum A in Fig.~\ref{f2}(c) indicates characteristic features typically observed in relatively small twist-angle TBGs: an R$^\prime$-band appeared at the G-band shoulder\cite{He2013a} as shown in Fig.~\ref{f2}(d-1) and the 2D peak width increased compared to the monolayer graphene.\cite{Kim2012} The R$^\prime$-band is attributed to intralayer electron-phonon (LO phonon) scattering process induced by moir{\'e} potentials.\cite{Eliel2018} Its position, which depends on the excitation laser energy (= 2.33 eV) and twist-angle, at 1605 cm$^{-1}$ is in good agreement with the reported calculation and the experiments.\cite{Eliel2018} One may consider that this peak is similar to the disorder-induced D$^\prime$-band caused by the intravalley phonon scattering process reported at 1620cm$^{-1}$.\cite{Ferreira2010} However, it is less possible to occur in our TBG as almost no signature of the D$^\prime$-band is seen in B area (see Fig.~\ref{f2}(d-2)), which includes a similar D-band intensity. The wider 2D peak is  due to multiple paths in the Raman transition processes due to modified electronic states.\cite{Kim2012, Coh}

\begin{figure}[tb]
\centering
\includegraphics[width=0.7\textwidth]{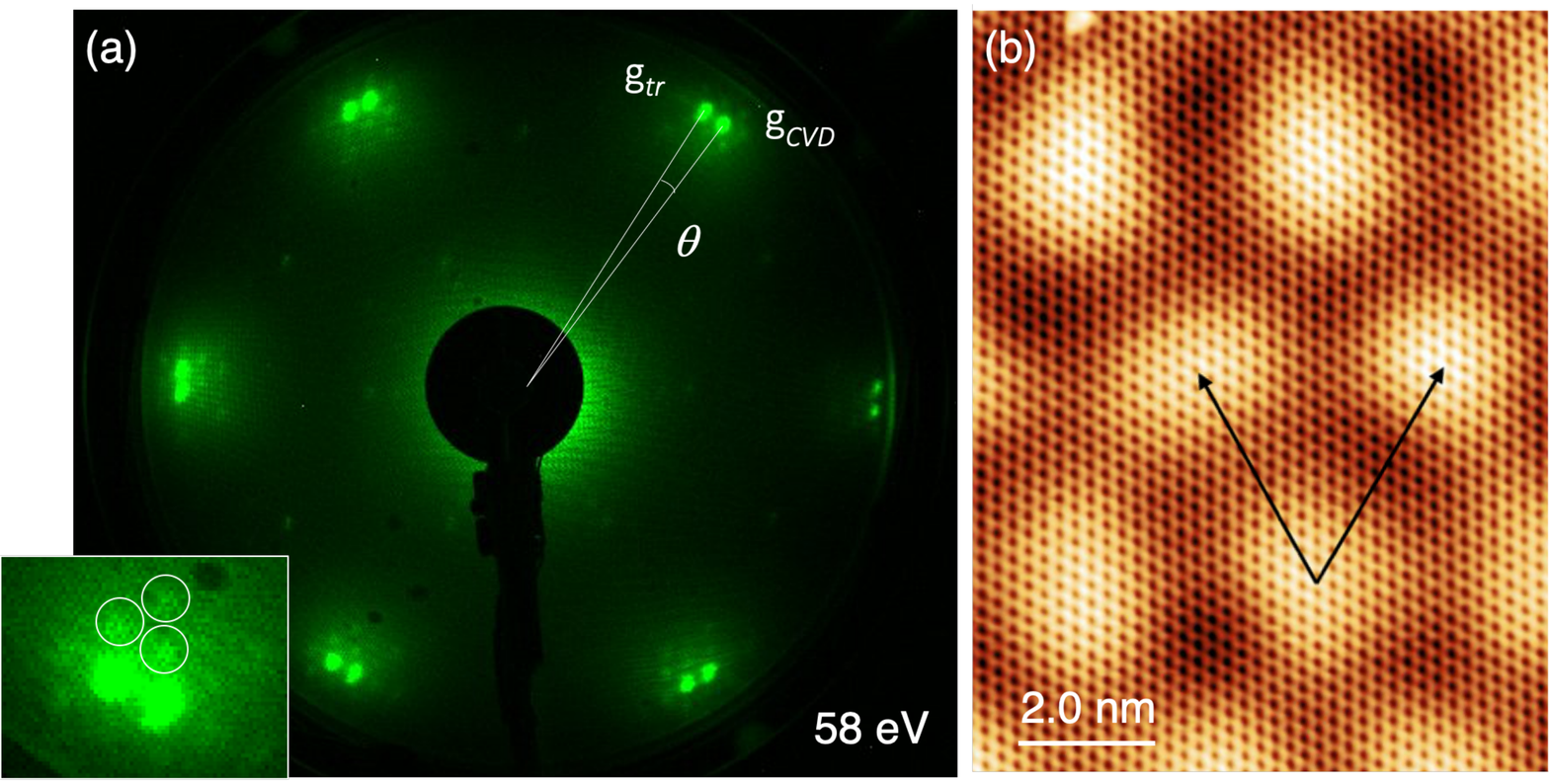}
\caption{(a) A LEED pattern of the sample containing TBG area. Two sets of 6-fold symmetric graphene diffraction spots from the bottom (g$_{CVD}$) and top (g$_{tr}$) layers are visible. Inset shows the magnified image of the graphene diffraction with satellite spots (circle). (b) An STM image taken at TBG area, showing graphene and moir{\'e} lattices.The arrows indicate an unit cell of the moir{\'e} lattice.The image was taken at the sample bias voltage= 0.8$~V$ and tunnnel current = 0.5$~nA$.}
\label{f3}
\end{figure}
Figure~\ref{f3}(a) shows a LEED pattern of the sample containing the TBG area. Two sets of graphene spots (bright spots having the 6-fold symmetry) and surrounding satellites (see inset) in addition to the faint spots of SiC substrate and the $(3\times3)$ buffer layer are seen. Each set of 6-fold pattern corresponds to CVD-grown (g$_{CVD}$) and transferred (g$_{tr}$) graphene. The relative angle between two sets of graphene spots should give a twist-angle, however, due to LEED optics imperfect alignment and screen mesh image it may includes fairly large errors. The satellite spots originated from moir{\'e} structure are thus more useful to estimate the twist-angle $\theta$, which is calculated from measured periodicity $L=3.7\pm0.4$~nm of the moir{\'e} pattern and the equation, $L=\frac{a}{2sin(\frac{\theta}{2})}$, where $a$ is a lattice constant of graphene.\cite{Beyer} The twist-angle $\theta$ is obtained to be $3.9\pm0.4^{\circ}$. To confirm the moir{\'e} periodicity scanning tunneling microscopy (STM) was conducted on the TBG area. Figure~\ref{f3}(b) clearly shows graphene and moir\'e lattices. The arrows indicate unit vectors of moir\'e cell, whose length is $\sim3.55$~nm. This is in good agreement with the LEED result. In the present experiment, the twist-angle can be controlled only within a few degree. More precise control and measurement of TBG twist-angle using other techniques, however, is necessary, especially in the case of lower twist-angle TBGs. Such equipment and procedures are currently under development.

\begin{figure}[tb]
\centering
\includegraphics[width=1\textwidth]{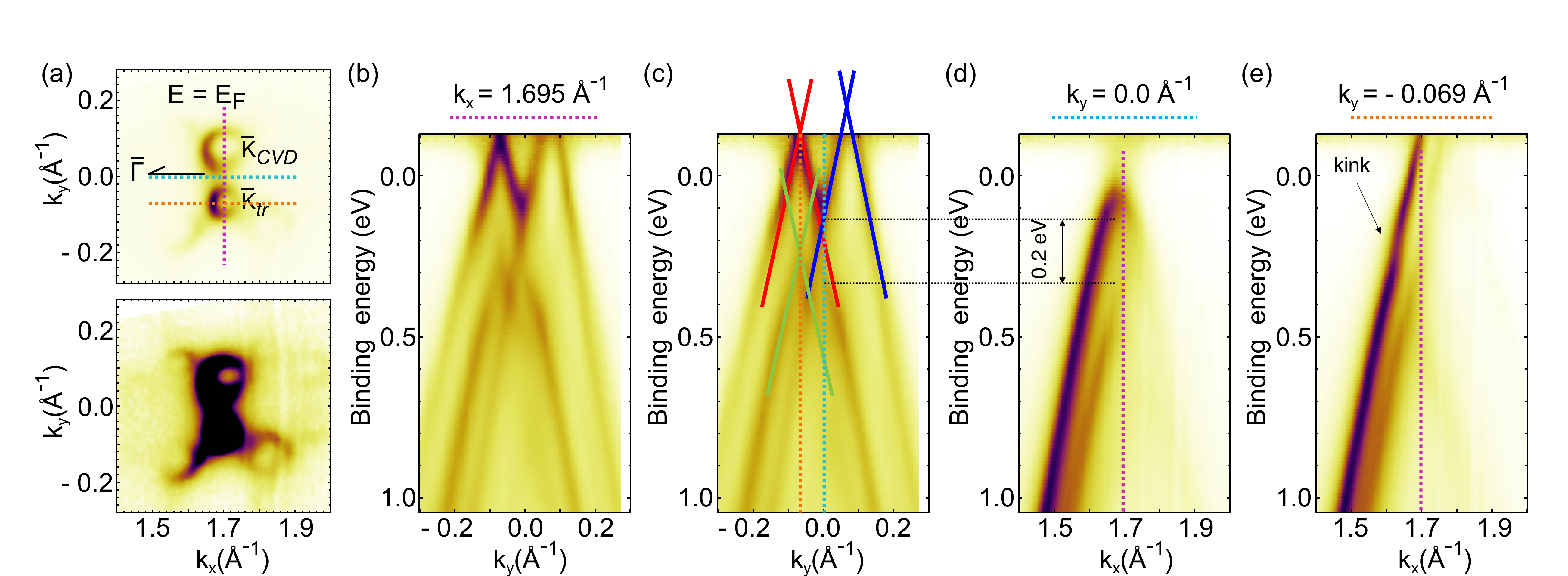}
\caption{ARPES constant energy maps and band dispersion. The k$_{\textrm{x}}$-axis is on the line between the middle of the two $\overline{\rm K}$ points of the graphene layers ($\overline{\rm K}_{CVD}$, $\overline{\rm K}_{tr}$) and $\overline{\Gamma}$ point. (a) Constant energy maps at E$_{\rm F}$ without (upper) and with (lower) image enhancement. Replica bands are seen at this energy. (b, c) Band dispersion at k$_{x} = 1.695$~\AA{}$^{-1}$ along the magenta dotted line in (a). Three Dirac bands are schematically shown as red, green and blue solid lines on the band map in (c). The $\overline{\rm K}$ points of the transferred graphene bands on the $(3\times3)$ (red) and 6R3 (green) substrates, and that of the CVD graphene (blue) on the $(3\times3)$ (red) substrate are at k$_{y} = -0.069$~\AA{}$^{-1}$ and 0.069~\AA{}$^{-1}$, respectively. The ARPES intensiy from the CVD graphene on the 6R3 substrate is weak.  (d) Band dispersion at k$_{y} = 0.0$~\AA{}$^{-1}$ along the cyan dotted line in (a). (e) Band dispersion at k$_{y} = - 0.069$~\AA{}$^{-1}$ along the orange dotted line in (a).}
\label{f4}
\end{figure}

Band structure was finally studied by ARPES using p-polarized synchrotron light  with a photon energy of 52 eV(KEK-PF, BL13) at room temperature.\cite{Toyoshima2013} Figure~\ref{f4}(a) shows the constant-energy band mapping at E = E$_{\rm F}$ around the two $\overline{\rm K}$ points ($\overline{\rm K}_{CVD}$, $\overline{\rm K}_{tr}$). Here, the observed ARPES intensity is normalized by the Fermi-Dirac distribution function at 300 K. Two Dirac cones derived from each sheet of TBG with several moir\'e replica bands are visible in the figures. One at positive k$_{y}$ originates from the bottom (CVD) layer, and the other at negative from the top (transferred) layer. This was confirmed by observing one Dirac cone in the band mapping at E$_{\rm F}$ for the area without the transferred layer. Figure~\ref{f4}(b) is band dispersion along k$_{y}$ at k$_{x}=1.695$~\AA{}$^{-1}$. There are two Dirac cones at k$_{x}=-0.069$~\AA{}$^{-1}$, and one at 0.069~\AA{}$^{-1}$, which are indicated as red, green and blue straight lines imposed over the band image in Fig.~\ref{f4}(c). It is known that monolayer graphene on 6R3 exhibits n-type doping\cite{Riedl2009}. However, graphene on $(3\times3)$ is of p-type. Thus, the observed n-type graphene (green cone) is identified as the top layer on the 6R3 substrate, and p-type graphene indicted by red (blue) lines corresponds to the top (bottom) layer of TBG on $(3\times3)$. The TBG bands on $(3\times3)$ are modulated at the binding energy E$_{\rm B}=+0.2$~eV where the blue and red Dirac cones intersect, and the band gap is open. Figure~\ref{f4}(d) shows a cross section of the band at k$_{y}= 0.0$~\AA{}$^{-1}$, where the blue and red Dirac cones intersect. The two bands are made by opening a band gap due to the interlayer coupling\cite{Ohta2012, Nishi2017}. The estimated value of the band gap at k$_{x} = 1.695$~\AA{}$^{-1}$ is about 0.2~eV. Figure~\ref{f4}(e) shows a cross section of the bands at k$_{y}=-0.069$~\AA{}$^{-1}$ (Fig.~\ref{f4}(a)). The Dirac band of the top layer is modulated around E$_{\rm B} = 0.2$ eV because of the interlayer coupling. Advantage of the present direct transfer method is demonstrated by the observed band modulations and observed several replica bands due to the strong interlayer coupling only possible with clean enough interface between graphene layers.

In summary, TBG was fabricated by directly bonding monolayer graphenes grown by the oxygen-added CVD method in a high vacuum. Easily exfoliating graphene is essential to conduct direct bonding with no use of any transfer assisting medium. Resulting TBG area was of sub-millimeter size, which enabled us to perform macro-probe analyses such as LEED and ARPES. The LEED pattern showed two sets of graphene diffraction spots rotated relative to each other and more importantly moir{\'e} superstructure diffraction. The moir{\'e} periodicity obtained by LEED and STM can help to estimate the twist-angle of $\sim3.9^{\circ}$. The ARPES spectra from the TBG near the K-points visualized band modifications at the intersection of Dirac cones and replica bands, due to strong interlayer coupling. These results confirm that our direct bonding method to fabricate TBG is promising to achieve a sizable, millimeter-order area TBG with a clean interface. The precise control of a twist-angle is, however, still remains challenging.

This work was supported by JSPS KAKENHI Grant Number JP19H02602 and 18H01146. ARPES measurements were performed under the approval of the Photon Factory Advisory Committee (Proposal No.2017G575).

\end{document}